\documentclass[journal]{IEEEtran}
\usepackage{graphicx}
\usepackage{float}
\usepackage{url}
\begin{document}
\title{Scanning Test System Prototype of p/sFEB for the\\
ATLAS Phase-I sTGC Trigger Upgrade}
%
%

\author{Xinxin Wang, Feng Li, Shengquan Liu, Peng Miao, Zhilei Zhang, Tianru Geng, Shuang Zhou,and Ge Jin
\thanks{Manuscript received Feb 9, 2018. This work was supported by the National Natural Science Foundation of China under Grants 11461141010 and 11375179, and in part by ``the Fundamental Research Funds for the Central Universities" under grant No. WK2360000005.}
\thanks{Xinxin Wang, Feng Li, Shengquan Liu, Peng Miao, Zhilei Zhang, Tianru Geng, Shuang Zhou, and Ge Jin are with State Key Laboratory of Particle Detection and Electronics, University of Science and Technology of China, Hefei 230026, P.R. of China (phone: +86-18019946174; e-mail: wxx10@mail.ustc.edu.cn,phonelee@ustc.edu.cn, lsqlsq@mail.ustc.edu.cn, mpmp@mail.ustc.edu.cn, zzlei@mail.ustc.edu.cn, gudujian@mail.ustc.edu.cn, neo@mail.ustc.edu.cn, goldjin@ustc.edu.cn ).}%
}

\maketitle
\thispagestyle{empty}

\begin{abstract}
the Pad Front End Board (pFEB) and the Strip Front End Board (sFEB) are developed for the ATLAS Phase-I sTGC Trigger Upgrade. The pFEB is used to to gather and analyze pads trigger, and the sFEB is developed to accept the pad trigger to define the regions-of-interest for strips readout. The performance of p/sFEBs must be confirmed before they are mounted on the sTGC detector. We will present the scanning test system prototype which is designed according to the test requirements of the p/sFEB. In this test system prototype, a simulation signal board is developed to generate different types of signal to the p/sFEB. PC software and FPGA XADC cooperate to achieve the scan test of analog parameter.
\end{abstract}


\section{Introduction}
%
%
%
%
\IEEEPARstart{A}{TLAS} \cite{1748-0221-3-08-S08003} detector which is one of the four experiments at Large Hadron Collider will fulfill Phase-I upgrade to extend the frontier of particle physics. The upgrade is going to replace the inner detector (Small Wheel, SW) \cite{collaboration2013new} of the end-cap muon spectrometer with the ¡°new small wheel¡± detector (NSW), which consists of the Small-strip Thin Gap Chamber (sTGC) \cite{mikenberg1988thin} and Micromegas (MM). The small-strip TGC (sTGC) in which the strip pitch is much smaller than that of the current ATLAS TGC will be applied for the NSW upgrade.

STGC contains pad, wire and strip readout. The pads are used to identify muon tracks roughly pointing to the interaction point (IP) through a 3-out-of-4 coincidence and define which strips need to be readout to obtain a precise measurement in the bending coordinate for the event selection. The Pad Front End Board (pFEB) \cite{li2018performance} is developed to readout pads signal to gather and analyze pads trigger. The Strip Front End Board (sFEB) \cite{li2016study} is developed to accept the pad trigger to define the regions-of-interest for strips readout. Both of pFEB and sFEB receive sTGC signals through the VMM3 \cite{Geronimo2013VMM1},\cite{VMM3} ASIC which handles 64 input signals, and outputs the trigger data and raw data of hit events.

About 2000 p/sFEBs will be produced for final delivery and engineering backup. Before the p/sFEB are mounted on the detector, we need to confirm the performance of all the p/sFEBs. According to the function of p/sFEB in the whole system, the performance testing of each p/sFEB includes baseline test, threshold DAC calibration, internal test pulse DAC calibration, gain test and dead channel test, each of which are very important for the front-end electronic system.

We develop the scanning test system prototype of the p/sFEB. In this test system prototype, a simulation signal board is developed to generate different types of signal to the p/sFEB. PC software and FPGA XADC cooperate to achieve the scan test of analog parameter. The PC software is written based on Qt platform using the standard C++.

\section{Function and Implementation}


\subsection{Test System Prototype Hardware Structure}

\begin{figure}[!t]
\centering
\includegraphics[width=3.5in]{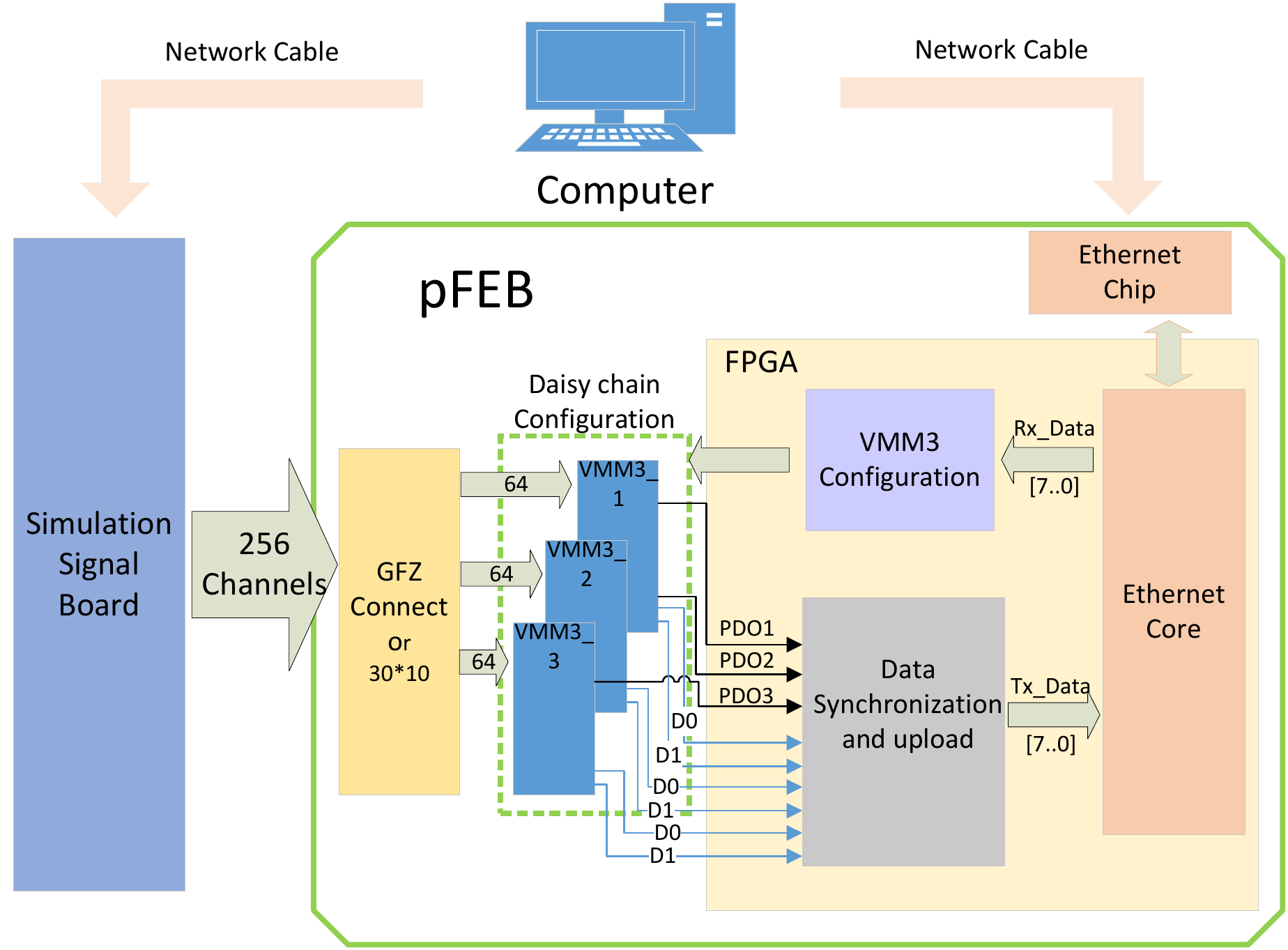}
\caption{Block diagram of Test Prototype System}
\label{fig1:}
\end{figure}
Fig.1 is a block diagram of the test system prototype. The p/sFEB includes three/eight VMM3 chips, one Kintex-7 FPGA for buffering VMM3 data, one Gigabit Ethernet Transceiver (GET), and connectors.The VMM3, which consists of 64 linear front-end channels, is an Application Specific Integrated Circuit (ASIC) for the detector. When the p/sFEB is connected to the sTGC detector, the analog signals from the sTGC detector are transmitted through the GFZ connector (10 * 30) to the p/sFEB and then into the three VMM3 chips via the protection circuit.

The simulation signal board \cite{hu2015note} can be used to provide p/sFEB with 256/512 test pulses via the GFZ connector. In addition, the VMM3 chip can generate an adjustable amplitude test pulse signal internally. The pulse signal is sent to each linear front-end channel of the VMM3. The VMM3 chip outputs digital signal into the FPGA, and the FPGA completes the corresponding readout and analysis works. The FPGA communicates with a computer through the network cable, achieving the VMM3 initialization and data transmission.

\subsection{Test System Prototype Implementation}

Each VMM3 chip requires 1728 configuration bits containing polarity, gain, peak time, threshold and other settings. Fig.2 shows the Qt-based GUI of the test system. The Qt-based GUI can complete the interaction between the computer and test board, automatic modification of parameters, issued commands and data acquisition. Pcap library is used to get access to Ethernet. In order to make the software run more smoothly and solve the problem of GUI stuck, the software uses a multi-threaded framework so that the data acquisition and user interface are built into different threads. Furthermore, this Qt-based GUI has the function of real-time data collection, analysis and display. This function is very important for p/sFEBs.
\begin{figure}[!t]
\centering
\includegraphics[width=3.5in]{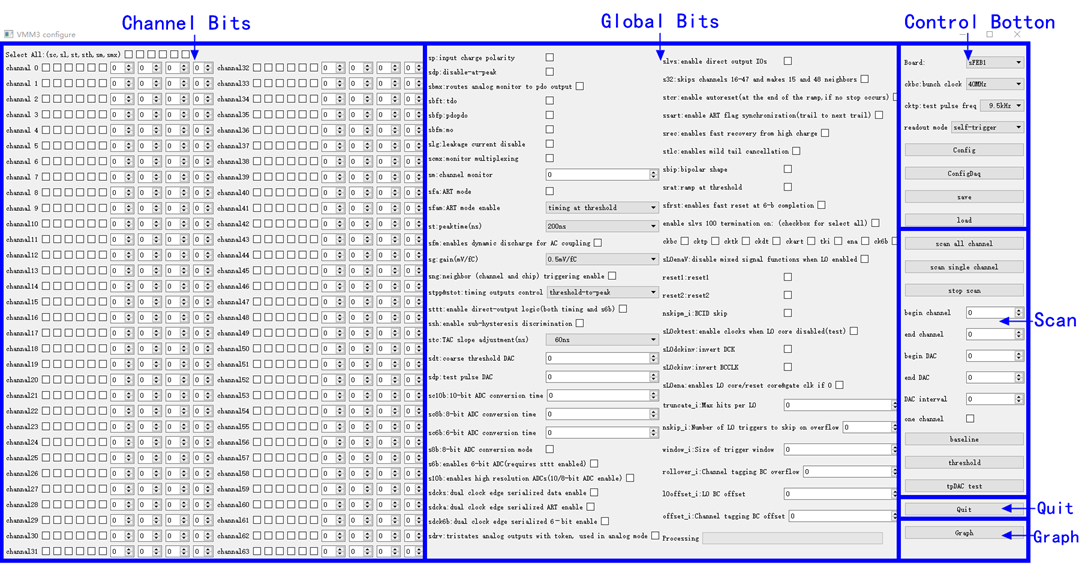}
\caption{GUI of Test System Prototype}
\label{fig2}
\end{figure}

It is noteworthy that the outputs of 192/512 channels baseline test, threshold DAC calibration test, and internal test pulse DAC calibration test are analog signals. Without this automatic test system, an oscilloscope needs to be connected to the p/sFEB to read the analog value. Using an oscilloscope to test is very time-consuming and not easy to take multiple measurements on average due to the need of testing many channels. So our system uses XADC in FPGA for automatic scanning to complete these three tests, which can change the configuration bits and sampling by XADC automatically. Finally, the digital data outputs to the computer through the Ethernet, and then achieving the purpose of rapid measurement.

\section{Test Results}

Fig.3 is the construction of the test platform. After many tests, it is proved that the test system is reliable and can truly reflect the relevant characteristics of p/sFEB. We develop the simulation signal board which can outputs 256 channels simulation signals in 6 kinds of mode to provide p/sFEB with 256/512 test pulses via the GFZ connector.
\begin{figure}[!t]
\centering
\includegraphics[width=3.5in]{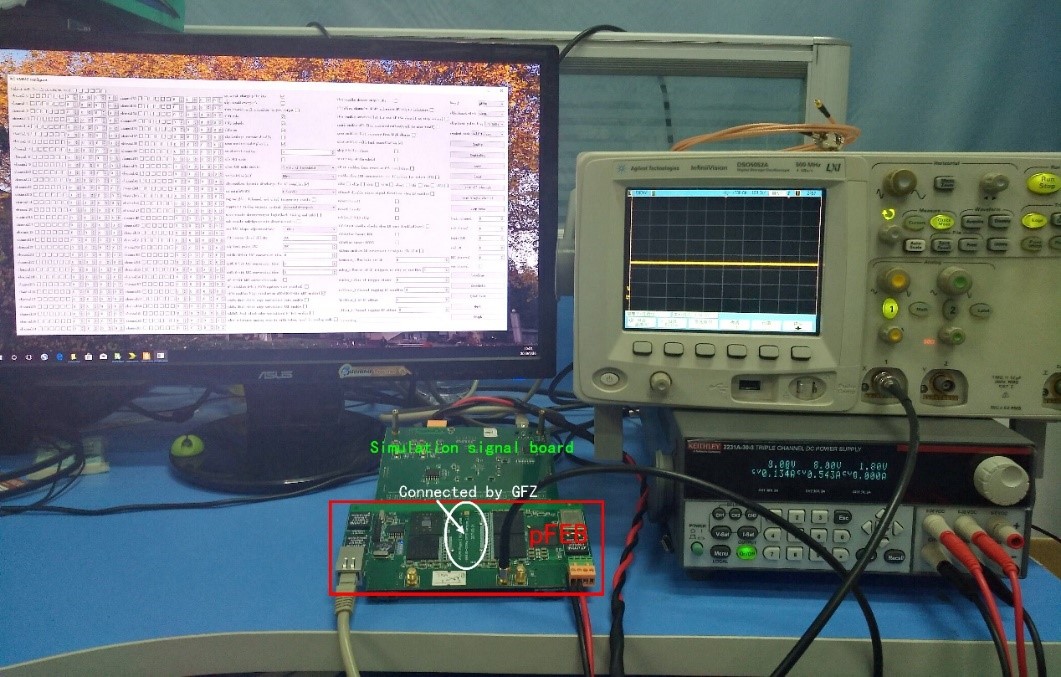}
\caption{the Construction of Test System Prototype}
\label{fig3}
\end{figure}

The Qt-based GUI will collect the hit event raw data and decode them. Then display the corresponding information. Fig. 4 shows the channel and amplitude test results for all the three VMM3s. In this GUI, the top three graphs displays hit event counts of VMM3 from channel 1 to channel 64, and the bottom three graphs display hits amplitude distribute information of one channel.We can find dead channels of each VMM3 from these graphs, and then further confirm this through another method to make the result more exactly.
\begin{figure}[!t]
\centering
\includegraphics[width=3.5in]{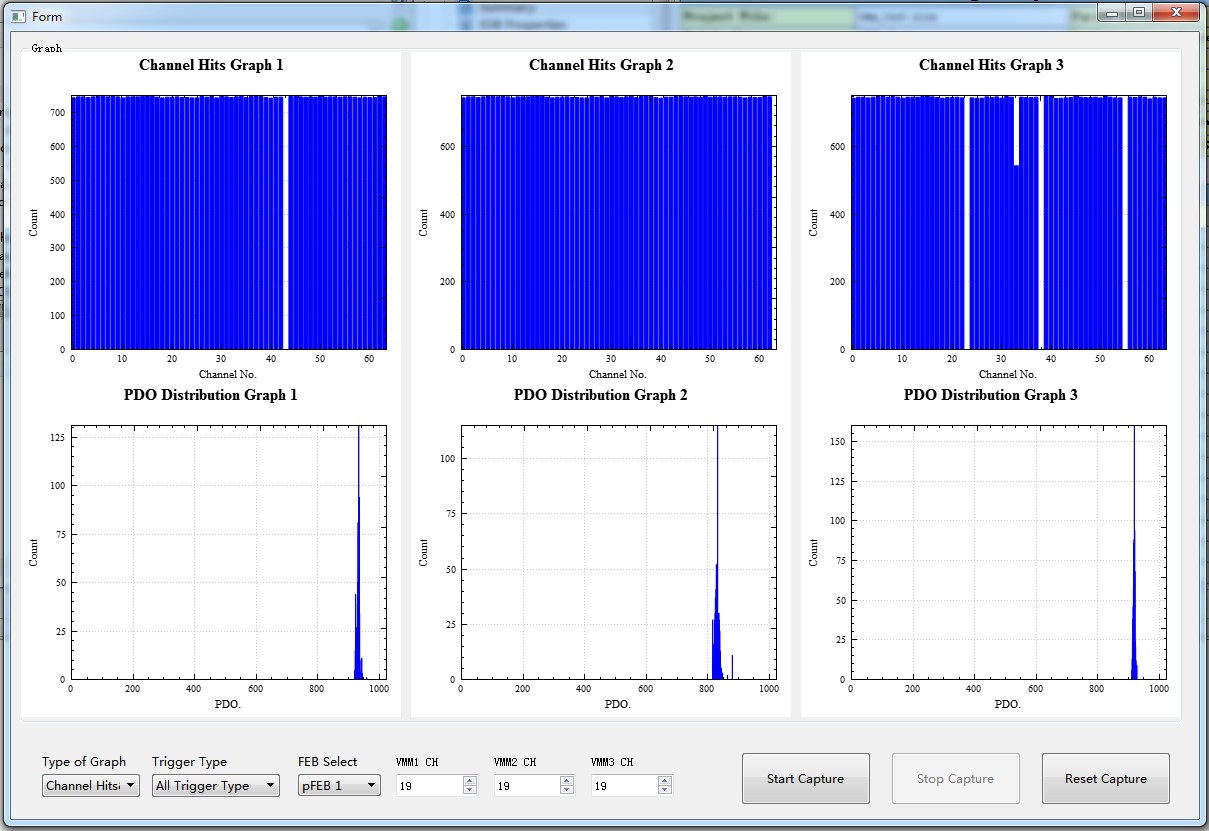}
\caption{the real-time result information  of VMM3}
\label{fig4}
\end{figure}

For the measurement of analog signals, we build the automatic scan test framework using some auxiliary analog inputs of the FPGA XADC. In this way, the analog signals can be measured several times to get the average value, which can increase the reliability and accuracy of the test result. Fig.5 shows the result of baseline scan test of a VMM3 chip. Each channel tests 100 times. From this graph, we can get the consistency and variability of the 64 channel baseline so that we can set the threshold of VMM3 chip.
\begin{figure}[!t]
\centering
\includegraphics[width=3.5in]{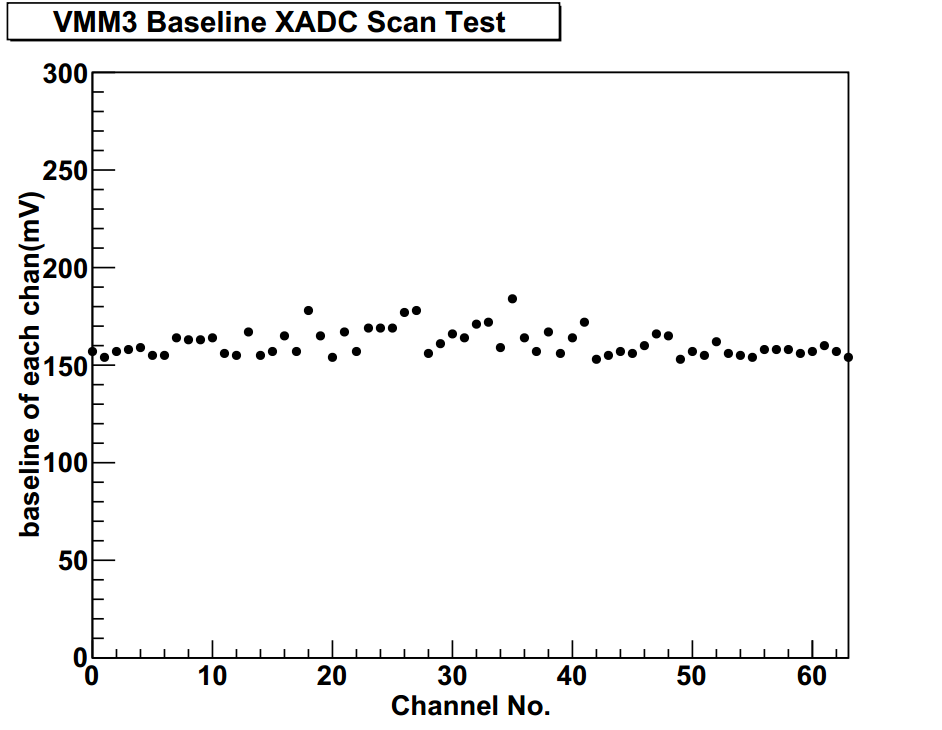}
\caption{Baseline scan result for 64 channels of the VMM3 chip}
\label{fig5}
\end{figure}
%
%

\section{Conclusion}

In this paper, we described the scanning test system prototype for p/sFEB in detail. In this test system prototype, a simulation signal board is developed to generate different types
of signals to the p/sFEB. PC software and FPGA XADC cooperate to achieve the scan test of analog parameter. With this system, we can test the circuit board quickly and reliably. In the early upgrade, the scanning test system is used for the p/sFEB performance test.

\newpage


\begin{thebibliography}{1}
\providecommand{\url}[1]{#1}
\csname url@samestyle\endcsname
\providecommand{\newblock}{\relax}
\providecommand{\bibinfo}[2]{#2}
\providecommand{\BIBentrySTDinterwordspacing}{\spaceskip=0pt\relax}
\providecommand{\BIBentryALTinterwordstretchfactor}{4}
\providecommand{\BIBentryALTinterwordspacing}{\spaceskip=\fontdimen2\font plus
\BIBentryALTinterwordstretchfactor\fontdimen3\font minus
  \fontdimen4\font\relax}
\providecommand{\BIBforeignlanguage}[2]{{%
\expandafter\ifx\csname l@#1\endcsname\relax
\typeout{** WARNING: IEEEtran.bst: No hyphenation pattern has been}%
\typeout{** loaded for the language `#1'. Using the pattern for}%
\typeout{** the default language instead.}%
\else
\language=\csname l@#1\endcsname
\fi
#2}}
\providecommand{\BIBdecl}{\relax}
\BIBdecl

\bibitem{1748-0221-3-08-S08003}
\BIBentryALTinterwordspacing
A.~Collaboration, ``The atlas experiment at the cern large hadron collider,''
  \emph{Journal of Instrumentation}, vol.~3, no.~08, p. S08003, 2008. [Online].
  Available: \url{http://stacks.iop.org/1748-0221/3/i=08/a=S08003}
\BIBentrySTDinterwordspacing

\bibitem{collaboration2013new}
A.~M. Collaboration, ``New small wheel technical design report,'' \emph{Tech.
  Des. Rep. CERNLHCC-2013-006, CERN}, 2013.

\bibitem{mikenberg1988thin}
G.~Mikenberg, ``Thin-gap gas chambers for hadronic calorimetry,'' \emph{Nuclear
  Instruments and Methods in Physics Research Section A: Accelerators,
  Spectrometers, Detectors and Associated Equipment}, vol. 265, no. 1-2, pp.
  223--227, 1988.

\bibitem{li2018performance}
F.~Li, S.~Liu, K.~Hu, X.~Wang, H.~Lu, X.~Wang, H.~Yang, T.~Geng, P.~Miao, and
  G.~Jin, ``Performance of pad front-end board for small-strip thin gap chamber
  with cosmic ray muons,'' \emph{IEEE Transactions on Nuclear Science},
  vol.~65, no.~1, pp. 597--603, 2018.

\bibitem{li2016study}
F.~Li, K.~Hu, X.~Wang, H.~Lu, T.~Geng, X.~Wang, H.~Yang, S.~Liu, L.~Han, and
  G.~Jin, ``The study of strip readout prototype for atlas phase-i muon trigger
  upgrade,'' in \emph{Real Time Conference (RT), 2016 IEEE-NPSS}.\hskip 1em
  plus 0.5em minus 0.4em\relax IEEE, 2016, pp. 1--3.

\bibitem{Geronimo2013VMM1}
G.~D. Geronimo, J.~Fried, S.~Li, J.~Metcalfe, N.~Nambiar, E.~Vernon, and
  V.~Polychronakos, ``Vmm1¡ªan asic for micropattern detectors,'' \emph{IEEE
  Transactions on Nuclear Science}, vol.~60, no.~3, pp. 2314--2321, 2013.

\bibitem{VMM3}
\BIBentryALTinterwordspacing
BNL, ``Atlas nsw electronics specifications component: Vmm,'' 2016. [Online].
  Available:
  \url{https://twiki.cern.ch/twiki/pub/Atlas/NSWelectronics/vmm33.pdf}
\BIBentrySTDinterwordspacing

\bibitem{hu2015note}
K.~Hu, H.~Lu, X.~Wang, F.~Li, F.~Liang, and G.~Jin, ``Note: The design of thin
  gap chamber simulation signal source based on field programmable gate
  array,'' \emph{Review of Scientific Instruments}, vol.~86, no.~1, p. 016116,
  2015.

\end{thebibliography}
\end{document}